\documentclass[prb,a4paper,twocolumn,showpacs,superscriptaddress]{revtex4}
\usepackage{amsmath}
\usepackage{amsfonts}
\usepackage{amssymb}
\usepackage{bm}
\usepackage{graphicx}

\begin{document}

\title{Tunneling of Cooper Pairs across Voltage Biased Asymmetric Single-Cooper-Pair Transistors}
\author{J. Lepp\"akangas}
\email[Electronic adress: ]{juha.leppakangas@oulu.fi}
\affiliation{Department of Physical Sciences,
P.O.Box 3000, FI-90014 University of Oulu, Finland}
\author{E. Thuneberg}
\affiliation{Department of Physical Sciences,
P.O.Box 3000, FI-90014 University of Oulu, Finland}
\author{R. Lindell}
\affiliation{Low Temperature Laboratory, Helsinki University of Technology, FI-02015 TKK, Finland}
\author{P. Hakonen}
\affiliation{Low Temperature Laboratory, Helsinki University of Technology, FI-02015 TKK, Finland}

\date{\today}


\begin{abstract}

We analyze tunneling of Cooper pairs across voltage biased asymmetric single-Cooper-pair transistors.
Also tunneling of Cooper pairs across two capacitively coupled Cooper-pair boxes is considered,
when the capacitive coupling and Cooper pair tunneling are provided by a small Josephson junction between the islands.
The theoretical analysis is done at subgap voltages, where the current-voltage characteristics depend strongly on
the macroscopic eigenstates of the island(s) and their coupling to the dissipative environment.
As the environment we use an impedance which
satisfies $\textrm{Re}[Z(\omega)]\ll R_Q$ and a few $LC$-oscillators in series with $Z(\omega)$.
The numerically calculated $I-V$ curves are compared with experiments where the
quantum states of mesoscopic SQUIDs are probed with inelastic Cooper pair tunneling.
The main features of the observed $I-V$ data are reproduced. Especially, we find traces of band structure in the higher excited states
of the Cooper-pair boxes as well as traces of multiphoton processes
between two Cooper-pair boxes in the regime of large Josephson coupling $E_J\gg E_C$.
\end{abstract}


\maketitle


\section{Introduction}\label{introduction}

A voltage biased small Josephson junction (JJ) has been shown to be a good probe of mesoscopic physics.
In recent years it has been used, for example, in the detection of resonances
in the electromagnetic environment~\cite{ingold1,ingold2} and noise spectroscopy~\cite{nongauss1,nongauss2}.
The theory of inelastic tunneling, known as the $P(E)$-theory, describes $I-V$ characteristics
resulting from incoherent tunneling of Cooper pairs, or quasiparticles, across the small JJ and simultaneous energy exchange between the
tunneling particle and its electromagnetic environment,
which is described by a set of $LC$-oscillators.
The standard $P(E)$-theory cannot, however, be used in the case of a non-Gaussian or anharmonic environment.
In this paper, a suitable model will be constructed to account for the anharmonicity of the environment.

This paper gives a quantum description for a system which is designed to
probe the excited states of a Cooper-pair box (CPB), or coupled boxes, by a small JJ.
We model the quantum evolution of a voltage biased asymmetric single-Cooper-pair transistor (SCPT) or a circuit consisting of three JJs in series with a
small middle JJ. The idea is, as in the $P(E)$-theory,
that the small JJ is probing the eigenstates of the CPBs,
which are then seen as current peaks at certain voltages.
This is possible since under the voltage bias well above the supercurrent peak, but still at subgap region,
the tunneling of a single Cooper pair across the small JJ is possible (nonvirtually) only if the environment
is able to absorb the energy $2eV$ released in the tunneling.

The environment of the small JJ consist of a CPB and a continuous spectrum of $LC$-oscillators describing dissipative
quantum mechanics induced by high frequency resistive properties of the leads and possible spurious
resonators in the transmission line or materials nearby the island.
In resonant situations the dynamics involve both excitation and relaxation of the CPB eigenstates and
one is, in principle, able to get information of both the energies as well as the relaxation times of the excited states.

Experimentally, the spectroscopy of the eigenstates using a small JJ as a probe have been done by Lindell
{\it et.\ al.\ }and the results are reported in Refs.~\onlinecite{rene1,rene2,rene3}.
In this set of experiments, traces of excited states, their anharmonicity and expected band structure were found from
the measured $I-V$ characteristics.
However, several unexplained phenomena seen there were the main motivations for
writing this more detailed description for the system.
We show that indeed the main features of the $I-V$ data can be explained by the
quantum mechanics of asymmetric SCPTs or coupled CPBs. The model explains, for example, the widening of the $I-V$ resonances
as result of a band structure of (coupled) CPBs and nonconstant peak splitting in the experiment
of Ref.~\onlinecite{rene1} as a result of multiphoton transitions between eigenstates of two CPBs.

The paper is organized as follows.
In section \ref{weak} we build a theory describing inelastic tunneling across the small JJ when it has an
anharmonic element, i.\ e.\ a CPB, in its environment.
In section \ref{strong} we discuss effects caused by slow relaxation and
section \ref{envi} is devoted to a quantitative discussion of the $I-V$ characteristics in
the case of a two CPB environment. Comparison between numerical calculations
and experiments is presented in section \ref{experiments}
and conclusions are given in section \ref{conclusion}.


\section{Incoherent Tunneling of Cooper Pairs across Asymmetric SCPT}\label{weak}

We model an asymmetric single-Cooper-pair transistor by taking the Josephson coupling across the
probe junction into account perturbatively.
The treatment describes incoherent tunneling of
Cooper pairs across the small JJ and simultaneous energy exchange between the tunneling Cooper pair, CPB
and the dissipative environment. It is valid if the relaxation rates of the excited states are higher than the
excitation rates induced by incoherent tunneling.

\begin{figure}[tb]
\begin{center}\leavevmode
\includegraphics[width=0.8\linewidth]{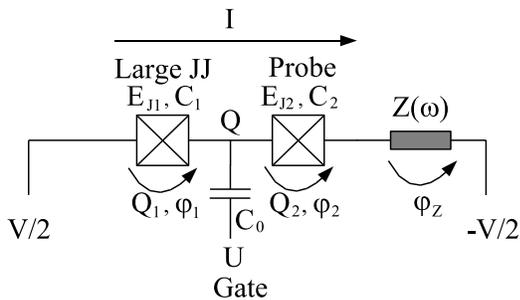}
\caption{A voltage biased SCPT.  We study the case $E_{J1}\gg E_{J2}$ and the smaller JJ is called the probe.
The superconducting leads (and the spurious environment of the island) are modelled through an impedance $Z(\omega)$.}
\label{kuvasset}
\end{center}
\end{figure}

A voltage biased SCPT is shown in Fig.~\ref{kuvasset}. We are interested in the case of strong asymmetry,
$E_{J1}\gg E_{J2}$.
The charging energy of the island, defined as $E_C=e^2/2C_{\Sigma}$ where $C_{\Sigma}=C_0+C_1+C_2$, can
however have an arbitrary value.
Disregarding the Cooper pair and quasiparticle tunneling across the probe
(the former is taken into account as a perturbation),
the Hamiltonian of {\it the environment of the probe} can be written formally as
\begin{gather}
H_{\rm env}=H_{\rm CPB}+H_{\rm EE}+H_{\rm int},
\label{mainn}
\end{gather}
where the CPB Hamiltonian is~\cite{band1,band2}
\begin{gather}
H_{\rm CPB}=\frac{\left(Q+Q_0\right)^2}{2C_{\Sigma}}-E_{J1}\cos(\varphi_1),
\label{box}
\end{gather}
and $Q_0=C_0U-(C_2+C_0/2)V$ is the quasicharge, $Q$ the island charge, a conjugated variable to the phase difference
$\varphi_1$. The Hamiltonian $H_{\textrm{EE}}$ models
dissipative electromagnetic environment and possible spurious fluctuators in the system.
Its characteristics are fully described by an impedance $Z(\omega)$, so formally it consist of an infinite number of harmonic oscillators.
The interaction term $H_{\rm int}$ describes linear coupling between the CPB and the electromagnetic environment
leading to dissipative quantum mechanics~\cite{dissipative}.

To account for the anharmonicity and the band structure of the CPB, we will proceed slightly differently than
is done in the $P(E)$-theory. The idea is to take the anharmonic parts into account separately. This is
possible in the limit $\textrm{Re}[Z(\omega)]\ll R_Q$,
where the interaction between the CPB and the dissipative environment is weak and one can use the Born-Markov approximation
when describing the evolution of the CPB under the Hamiltonian (\ref{mainn}).
In this limit the effect of the operator $H_{\rm EE}+H_{\rm int}$ for the
CPB can be described by a transformation $V\rightarrow V+V_f$ in Eq.~(\ref{box}),
where $V_f$ describes fluctuations from the average value $V$~\cite{zeite,brink}.
Therefore one can use an effective Hamiltonian for the CPB
\begin{gather}
H_{\rm BM}=H_{\rm CPB}-Q_{\rm{int}}V_f,
\end{gather}
where $Q_{\rm{int}}=C_2Q/C_{\Sigma}$ (we have assumed that $C_0\ll C_{\Sigma}$).
The autocorrelation function of the fluctuating voltage is related to the dissipative properties of the impedance $Z(\omega)$ via the
quantum fluctuation-dissipation theorem
\begin{eqnarray}
& & \langle V_f(t)V_f(0)\rangle_{\omega}=\int_{-\infty}^{\infty}dte^{i\omega t}\langle V_f(t)V_f(0)\rangle=\nonumber\\
& & 2\textrm{Re}[Z(\omega)]\frac{\hbar\omega}{1-\exp(-\hbar\omega/k_BT)}.
\label{fluct}
\end{eqnarray}
Since fluctuations are only a small perturbation to the CPB, their effect is to induce transitions between the
unperturbed states, i.\ e.\ the eigenstates of the CPB.
The transition rate between the eigenstates $\vert i\rangle$ and $\vert f\rangle$ is obtained by the
golden rule calculation
\begin{gather}
\gamma_{f\leftarrow i}=\frac{2\textrm{Re}[Z(\omega)]}{\hbar^2}\vert\langle f\vert Q_{\rm int}\vert i\rangle\vert^2\frac{E_{if}}{1-\exp(-E_{if}/k_BT)},
\label{rates2}
\end{gather}
where $E_{if}=E_i-E_f=\hbar\omega$ is the difference between the corresponding eigenenergies.

We proceed by noting that the rates (\ref{rates2}) define the lifetimes, and therefore
also the linewidths, of the energy levels in the Cooper-pair box.
The full width at half maximum (FWHM) $\Delta_{\alpha}$ of the state $\vert \alpha\rangle$ is then
\begin{gather}
\Delta_\alpha=\hbar\sum_f\gamma_{f\leftarrow \alpha},
\label{width}
\end{gather}
and the density of the excited states broadens from the sum of delta-functions to sum of Lorentzians, i.\ e.\ the density of states changes as
\begin{gather}
\sum_{\alpha}\delta(E-E_{\alpha})\rightarrow\sum_{\alpha}\frac{2}{\pi}\frac{\Delta_{\alpha}}{4(E_{\alpha}-E)^2+\Delta_{\alpha}^2}.
\label{trans4}
\end{gather}
Finally, we include the Josephson coupling $E_{J2}\cos(\varphi_2)$
describing tunneling of Cooper pairs across the probe junction and simultaneous excitations
of its environment as (another) perturbation.
Using $\varphi_{\Sigma}=\varphi_1+\varphi_2+\varphi_Z=2eVt/\hbar$, where $\varphi_Z$ is the phase difference across
the impedance, one finds the time dependent perturbations for the positive and negative direction tunneling
$M_{\pm}=E_{J2}\exp{ \left[\pm i\left(\varphi_1+\varphi_Z-2eVt/\hbar\right)\right] }/2$.

The transition rates due to perturbations $M_{\pm}$ between the eigenstates of the $H_{\rm{env}}$
are effectively described by transition rates between the states
$\vert\alpha\rangle\vert\rm{env}\rangle$, which consist of two independent parts: {\it broadened} CPB states
$\vert\alpha\rangle$ and continuous distribution of the environmental $LC$-oscillators,
whose free evolution is described by $H_{\rm EE}$ (Born approximation).
The $LC$-environment can be traced out similarly as in the $P(E)$-theory and the golden rule
transition rates between the CPB states $\vert i\rangle$ and $\vert f\rangle$ become then
\begin{gather}
\begin{split}
\Gamma_{f\leftarrow i}^{\pm}= & \frac{E^2_{J2}}{\hbar}\int_{-\infty}^{+\infty} dE' P(\pm 2eV-E')\vert\langle f\rvert e^{\pm i\varphi_1}\lvert i\rangle\vert^2\times\\
&\frac{\Delta_i+\Delta_f}{4(E_f-E_0-E')^2+(\Delta_i+\Delta_f)^2},
\end{split}
\label{trans4}
\end{gather}
where the $P(E)$-function
is the same as for a system consisting of a probe junction with a capacitance $C_{12}=(1/C_1+1/C_2)^{-1}$ in series with the
impedance $Z(\omega)$.
If $Z(\omega)$ is a constant $R$ ($\ll R_Q$), the main contribution of the
$P(E)$-function becomes from low energies where it
is approximately a Lorentzian with a linewidth $\Delta_{\rm{env}}=4\pi k_BTR/R_Q$ centered at $E=0$\cite{ingold2}.
Therefore it convolutes the original transition rates (\ref{trans4}) to
\begin{gather}
\begin{split}
\Gamma_{f\leftarrow i}^{\pm}= & \frac{E^2_{J2}}{\hbar}\vert\langle f\rvert e^{\pm i\varphi_1}\lvert i\rangle\vert^2\times\\
&\frac{\Delta^{\textrm{total}}_{if}}{4(E_f-E_i\mp 2eV)^2+(\Delta^{\textrm{total}}_{if})^2}.
\end{split}
\label{trans5}
\end{gather}
where $\Delta^{\textrm{total}}_{if}=\Delta_i+\Delta_f+\Delta_{\textrm{env}}$. We see that
there are two sources of broadening of the resonances: widening due to finite lifetimes of the CPB-eigenstates
($\Delta_{\alpha}$:s) and widening due to low frequency fluctuations of the $LC$-environment ($\Delta_{\rm{env}}$).
The separation of the CPB and its environment
holds also for the case where the environment has several modes which are nondegenerate with the CPB eigenstates.
The degeneration, or almost degeneration, would lead to similar splitting of the states as described in section \ref{envi}.

Each transition will occur simultaneously with a transfer of $2e$ of charge across the system and
the current is therefore
\begin{gather}
I=2e\sum_{fi}P_{i}\left(\Gamma_{f\leftarrow i}^+-\Gamma_{f\leftarrow i}^-\right),
\label{current}
\end{gather}
where the probabilities $P_{i}$ for occupancies of the CPB eigenstates are given by the canonical equilibrium distribution.
If $k_BT\ll E_{1}-E_{0}$ and $Z(\omega)=R$ then
\begin{gather}
\begin{split}
I(V)=&2e\sum_{f}\Gamma_{f\leftarrow 0}^+=\sum_{f}\frac{2eE^2_{J2}}{\hbar}\vert\langle f\rvert e^{i\varphi_1}\lvert 0\rangle\vert^2\times\\
&\frac{\Delta^{\textrm{total}}_f}{4(E_f-E_0-2eV)^2+(\Delta^{\textrm{total}}_f)^2}.
\label{tulos}
\end{split}
\end{gather}
One sees that $I-V$ peaks can be identified with energy levels of the environment~\cite{ingold2,holst},
which in this case is the Cooper-pair box.

We have verified that the $I-V$ characteristics obtained from Eq.~(\ref{tulos}) reduce to the ones obtained from the
$P(E)$-theory, if the larger JJ is described as an $LC$-oscillator.
However, Eq.~(\ref{tulos}) is also valid for JJ-environment with evident anharmonicity or band structure, and therefore
is not limited to the harmonic approximation.


\section{Effects due to Slow Relaxation}\label{strong}

For the expression ({\ref{trans5}}) to hold, it is vital that the system relaxes
rapidly to the ground state, since the golden rule calculation is justified only if the excitation rates of the CPB eigenstates,
induced by the probe,
are smaller than the relaxation times, caused by the CPB's coupling to the dissipative environment.
The irreversible interaction with the dissipative environment
has to "cut" the evolution to the excited state quickly, otherwise the tunneling across
the probe would turn from incoherent to coherent.
On the other hand, in the opposite case of very slow relaxation,
one would obtain Rabi oscillations between the CPB eigenstates $\vert 0\rangle$ and $\vert\alpha\rangle$
when initially starting from the state $\vert 0\rangle$ with $2eV=E_{\alpha}-E_0$.
This limit can also be analyzed in the Born-Markov approximation~\cite{lt24},
but generally the problem needs an analysis of the time evolution of the
whole density matrix and Markovian approximation cannot be used~\cite{zeno1}.

To obtain approximative results in all regions,
we use the model derived in section \ref{weak} with modified probabilities for occupations.
We redefine the diagonal elements of the density matrix by the ones obtained from the equilibrium master equation
\begin{gather}
\sum_{f\neq i}\left[P_{f}(\Gamma_{i\leftarrow f}+\gamma_{i\leftarrow f})-P_{i}(\Gamma_{f\leftarrow i}+\gamma_{f\leftarrow i})\right]=0,
\label{mestari}
\end{gather}
for each $i$. The $\gamma$:s are the relaxation rates caused by the fluctuating voltage
across the CPB, Eq.~(\ref{rates2}),
whereas the $\Gamma$:s are the rates induced by Cooper pair tunneling across the probe, Eq.~(\ref{trans5}).
The method assumes that all the sequential transitions are independent of each other, which is not always true. However,
the method reduces to the one considered in section \ref{weak} when the relaxation dominates
the excitation and, according to our numerical calculations, gives similar results {\it for the first order tunneling processes}
(single Cooper pair tunnels across the probe with simultaneous excitation of the CPB) even in the regime of very slow relaxation,
as long as the SCPT is highly asymmetric.
Therefore it is safe to assume that the first order processes are well approximated
by Eq.~(\ref{current}) with the equilibrium probabilities obtained from Eqs.~(\ref{mestari}).
Also, there is no experimental evidence of higher order resonances, which might be due to their weak
nature to be washed out by the so called Zeno effect~\cite{zeno1}.


\section{Two Capacitively Coupled CPBs}\label{envi}

To generalize the treatment of the preceding sections,
we do a perturbative treatment for three JJs in series, where the middle one acts as a probe.
The configuration can be seen to consist of two capacitively coupled Cooper pair boxes\cite{nakamura2003},
where the capacitive coupling is in parallel with a small tunneling element.
Since we use similar models for analyzing the experiments in section \ref{experiments},
we concentrate on the characteristics of this model a bit deeper.
We also note that one of the larger JJs
could as well be an $LC$-oscillator describing spurious resonance at frequency $\omega_p=1/\sqrt{LC}$ in the environment.
If the energy quantum $\hbar\omega_p$ is almost the same as any excitation energy $E_n-E_m$ between two relevant
eigenstates of the CPB, or the state is long living, it cannot be modelled by the $P(E)$-function in Eq.~(\ref{trans4}),
but the following treatment is valid.

\begin{figure}[tb]
\begin{center}\leavevmode
\includegraphics[width=0.85\linewidth]{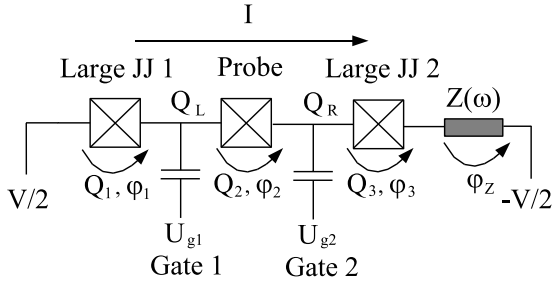}
\caption{Three Josephson junctions in series with the voltage source.
One of the large JJs could as well as be an $LC$-oscillator, modelling the effect of a spurious resonance.}
\label{manyjunctions}
\end{center}
\end{figure}

The system in consideration consists of three JJs in series connection with the voltage source
and the smallest junction is in the middle, Fig.~\ref{manyjunctions}.
Following the steps done in section \ref{weak}, we first neglect
the Josephson coupling energy of the probe and write down the Hamiltonian of two capacitively coupled Cooper pair boxes
\begin{gather}
\begin{split}
&H_{\rm 2CPB}=\frac{\left(Q_L+Q_0\right)^2}{2C_L}+\frac{\left(Q_R+Q_0'\right)^2}{2C_R}\\
&-E_{J1}\cos(\varphi_1)-E_{J3}\cos(\varphi_3)+\frac{C_{123}Q_LQ_R}{C_1C_3},
\end{split}
\label{CCPB}
\end{gather}
where $\varphi_1$ and $Q_L$ are conjugated variables
and similarly for $\varphi_3$ and $Q_R$. The capacitances of the islands are
(assuming that $C_{gi}\ll C_i$, where $C_{gi}$ is the capacitance of the gate $i$)
$C_L=C_1+C_{23}$ where $C_{23}=(1/C_2+1/C_3)^{-1}$,
$C_R=C_3+C_{12}$ and $C_{123}=(1/C_1+1/C_2+1/C_3)^{-1}$.
The quasicharges become then $Q_0\approx C_{g1}U_{g1}+C_{g2}U_{g2}C_2/C_{R}-C_2V$
and $Q_0'\approx -C_{g2}U_{g2}-C_{g1}U_{g1}C_2/C_{L}-C_2V$.
One sees that the Hamiltonian consist of two CPB Hamiltonians,
which are then coupled by the last term in Eq.~(\ref{CCPB}).
Similarly as in section \ref{weak}, we first calculate the linewidths of the eigenstates of the coupled system
(but now $Q_{\rm{int}}=Q_LC_2/C_{L}+Q_RC_2/C_{R}$), then
use the fact that the phase difference $\varphi_{\Sigma}$ is a classical variable and
take the tunneling across the probe into account perturbatively by
considering the broadened states of the coupled CPBs and the environmental states separately.
The current is obtained from Eq.~(\ref{current}), similarly as before.

The behaviour of the eigenstates and energies of the Hamiltonian (\ref{CCPB}) can be analyzed analytically in
the limit $E_{J1},E_{J3}\gg e^2/2C_L,e^2/2C_R$. For simplicity let us assume that $E_{J1}=E_{J3}=E_J$ and
$C_L=C_R\approx C_1=C_3=C$. Two ``splitting" effects contribute to the final energy level structure of the coupled CPBs.
First, in the harmonic approximation of Eq.~(\ref{CCPB}) the CPB:s behave as $LC$-oscillators.
The degeneracy of the identical $LC$-oscillators is removed by the interaction term
$C_{123}Q_LQ_R/C_1C_3\approx C_2Q_LQ_R/C^2$ (presuming that $C_2\ll C$), and
diagonalising the quadratic Hamiltonian one sees that the system behaves as it would consist of
two independent oscillators with the original inductances but with capacitances $\tilde C_{\pm}=C^2/(C\pm C_2)$.
For small $C_2/C$ this leads to mode frequencies $\omega_p\pm\omega_pC_2/2C$.
Secondly, if one takes into account the first nonharmonic terms $-E_{J}\varphi_i^4/4!$ in the
Hamiltonian (\ref{CCPB}) and sets $C_2Q_LQ_R/C^2=0$, one obtains also energy level splitting effects due to combined energy levels of
two anharmonic oscillators.
The energy levels of single CPBs become $E_1=\hbar\omega_p-E_C$, $E_2=2\hbar\omega_p-3E_C$, $E_3=3\hbar\omega_p-6E_C\ldots$.
New levels appear due to simultaneous excited states of the boxes
\begin{gather}
\vert 2^*\rangle=\vert 1\rangle\vert 1\rangle\\
\vert 3^*\rangle=\frac{1}{\sqrt{2}}(\vert 1\rangle\vert 2\rangle+\vert 2\rangle\vert 1\rangle)
\end{gather}
with the corresponding eigenenergies $E_{2^*}=2\hbar\omega_p-2E_C$ and $E_{3^*}=3\hbar\omega_p-4E_C$.
The energy level $2\hbar\omega_p$ therefore ``splits" into two nearby energy levels $E_2$ and $E_{2^*}$.

When both of the above effects are included, more mixing of the states is obtained.
The $n$:th excited state splits into $n+1$ states and, for example, the state $\vert 1\rangle$
splits into states (using the first order perturbation theory)
\begin{gather}
\vert 1s\rangle=\frac{1}{\sqrt{2}}(\vert 1\rangle\vert 0\rangle+\vert 0\rangle\vert 1\rangle)\\
\vert 1a\rangle=\frac{1}{\sqrt{2}}(\vert 1\rangle\vert 0\rangle-\vert 0\rangle\vert 1\rangle)
\label{anhar}
\end{gather}
with the eigenenergies $E_{1s}=\hbar\omega_p-z-E_C$ and $E_{1a}=\hbar\omega_p+z-E_C$, where $z=C_2\hbar\omega_p/2C$.
Similarly for the state $\vert 2\rangle$
\begin{gather}
\vert 2s\rangle=c_1(\vert 2\rangle\vert 0\rangle+\vert 0\rangle\vert 2\rangle+c^+\vert 1\rangle\vert 1\rangle)\\
\vert 2a\rangle=\frac{1}{\sqrt{2}}(\vert 2\rangle\vert 0\rangle-\vert 0\rangle\vert 2\rangle)\\
\vert 2^*\rangle=c_2(\vert 2\rangle\vert 0\rangle+\vert 0\rangle\vert 2\rangle+c^-\vert 1\rangle\vert 1\rangle),
\end{gather}
where $E_{2s}=2\hbar\omega_p-5E_C/2-z'$,
$E_{2a}=2\hbar\omega_p-3E_C$,
$E_{2^*}=2\hbar\omega_p-5E_C/2+z'$,
$c^{\pm}=(-E_C/2\pm z')/\sqrt{2}z$, $z'=\sqrt{E_C^2+16z^2}/2$ and $c_i$ are normalizing factors.
These states give both behaviours discussed in the preceding paragraph as the limiting cases of $E_C\rightarrow 0$ and $z\rightarrow 0$,
respectively. The states $\vert 1a\rangle$ and $\vert 2a\rangle$ do not lead to
current peaks since they contain antisymmetric combination of the states
and therefore the elements $\vert\langle f\vert\exp(\pm i(\varphi_L+\varphi_R))\vert 0\rangle\vert$ vanish.
We still pick up the energies of the states
$E_{3s}=3\hbar\omega_p-5E_C-z-\sqrt{E_C^2+4z^2-2E_Cz}$ and
$E_{3^*}=3\hbar\omega_p-5E_C-z+\sqrt{E_C^2+4z^2-2E_Cz}$.

For higher anharmonicity ($E_J\sim E_C$), where the band structure will become evident for the eigenstates
and the perturbative treatment of the cosine-potential is no longer valid,
we have to resort to numerical solution.
We calculate the eigenstates by diagonalising the Hamiltonian in a product basis
of two noninteracting CPBs, for given values of gate voltages (quasicharges). The lowest eigenstates can be obtained quite accurately from
an economical sized matrix equation, since the eigenstates are usually close to the states of this basis (due to small
capacitive coupling), justifying also the ``product state labelling" of the final states.
After the Hamiltonian is diagonalised (and the transition rates have been calculated) one has to solve
Eqs.~(\ref{current}) and (\ref{mestari}) for each value of $V$ to obtain the $I-V$ characteristics for given $U_{gi}$.

In section \ref{experiments2} we model experiments using a similar circuit but including also two extra $LC$-oscillators
in series with the three JJ system. The previous $I-V$ characteristics of the coupled CPBs are still preserved but multiphoton transitions
with the external $LC$-oscillators are also obtained, which is the motivation for this procedure. In practice, the
extra $LC$-oscillators can be modelled as JJs and the Hamiltonian of the system can be written as
\begin{gather}
H=\sum_{k,l}\frac{1}{2}(C^{-1})_{kl}q_kq_l-\sum_{i}E_{Ji}\cos(\varphi_i)
\label{multi1}
\end{gather}
where $q_k=Q_k-Q_{k+1}$, $Q_k$ is the charge gone through the $k$:th JJ, a conjugate variable to $\varphi_i$,
$C$ is a capacitance matrix\cite{ingold1}.
This Hamiltonian fully determines the energy bands, i.~e.~the ranges where peaks can occur in $I-V$ characteristics.
In order to determine further details such as the peak positions for given gate voltages,
one has to complete the Hamiltonian with linear terms in charge.
Such terms result from voltage sources and can be deduced from single tunneling events\cite{ingold1,tinkham}.
However, it turns out that in the experiments to be analyzed, the quasicharge is averaged over
all values and the resonance positions become immune to these terms.

The relaxation rates due to photon emission to the electromagnetic environment $Z(\omega)$ are also determined by
linear terms, through the fluctuating operator $Q_{\rm{int}}V_f=\sum_{i}a_iQ_iV_f$.
For the $LC$-oscillators this relaxation channel does not lead to observed rates and must be
enhanced by introducing resistors in parallel with $L$ and $C$. An analogous procedure is to take
the coefficients $a_{LC}$ as fitting parameters and use the operator $\sum_{i}a_{LCi}Q_{LCi}V_{LCi}$ as a perturbation,
where the fluctuations $V_{LCi}$ are uncorrelated but have the original properties.
For the large JJs $a_i$'s are theoretically defined by the ratios
$C_2/C_{L/R}$, as was seen in the beginning of this section. The ratios are locked when fitting the observed energy level structure of the coupled system.
But in real systems the capacitive coupling of CPBs can be effectively reduced by decoherence effects,
for example by thermal fluctuations or dissipation, leading to a decrease in the ``observed" $C_2$.
Also effects related to materials nearby the CPB islands seem to be able to increase relaxation\cite{nakamura2005}.
Therefore in modelling the relaxation rates of coupled CPBs, one is
forced to take the corresponding coefficients $a_i$ as independent fitting parameters.


\section{Comparison to Experiments}\label{experiments}

The $I-V$ characteristics of similar systems as discussed above were
measured experimentally by Lindell {\em et.\ al.\ }and reported in Refs.~\onlinecite{rene1,rene2,rene3}.
In these experiments, different kind of environments for the probe junction, consisting of one or several SQUIDs
and two or four leads, were used under different magnetic fields and gate voltages.
Using SQUIDs as the large JJs, the system could be studied {\em in situ}
from the harmonic behaviour ($E_{J1}\gg E_C$) to the region where the anharmonicity and band structure
become crucial ($E_{J1}\sim E_C$), by applying magnetic flux to the SQUID loops. This property also helped in the analysis of
the data, since resonances coming from the spurious environment did not react to the applied magnetic field,
at least not in the same way as the resonances coming from the CPBs.

\begin{figure}[tb]
\begin{center}\leavevmode
\includegraphics[width=\linewidth]{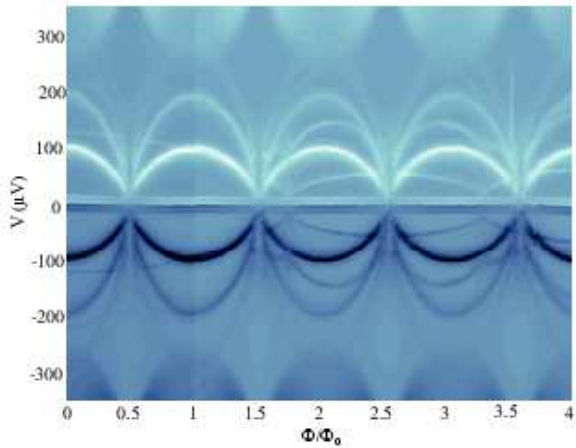}
\caption{
A 2D-surface plot of the measured probe junction current as a function of voltage and flux
through the SQUID loop. The resonances originating from the SQUIDs show $\Phi_0$-periodicity.
The figure also shows resonances of Fraunhofer/Airy type with weaker dependence on $\Phi$, and multiphoton transitions.
The $I-V$ characteristics of this 4-SQUID sample are analyzed more quantitatively in section \ref{experiments2}.
}\label{sjlc8}\end{center}\end{figure}

The probe junction current as a function of voltage and external flux is shown in Fig.\ \ref{sjlc8}
as a 2D-surface plot. The dominant current peaks show periodicity as a function of the flux $\Phi$ through
the SQUID loop with the period of the flux quantum $\Phi_0=h/2e$.
This clearly points that they are originating
from the SQUIDs and allows the identification of the different SQUID excitation states from the more
complicated $I-V$ structure due to the rest of the electromagnetic environment.
In addition to the periodic  structures due to the SQUID environment, one can see additional states with longer
and non-constant periods. It is likely that these are due to large, additional, Josephson junctions that are
created in the two-angle evaporation technique used to fabricate the sample.
The patterns have the Fraunhofer/Airy behavior as expected for a large Josephson junction that is penetrated by a
magnetic field\cite{tinkham}.

\subsection{The 1-SQUID Experiment}\label{experiments1}

\begin{figure}[tb]
\begin{center}\leavevmode
\includegraphics[width=0.80\linewidth,angle=-90]{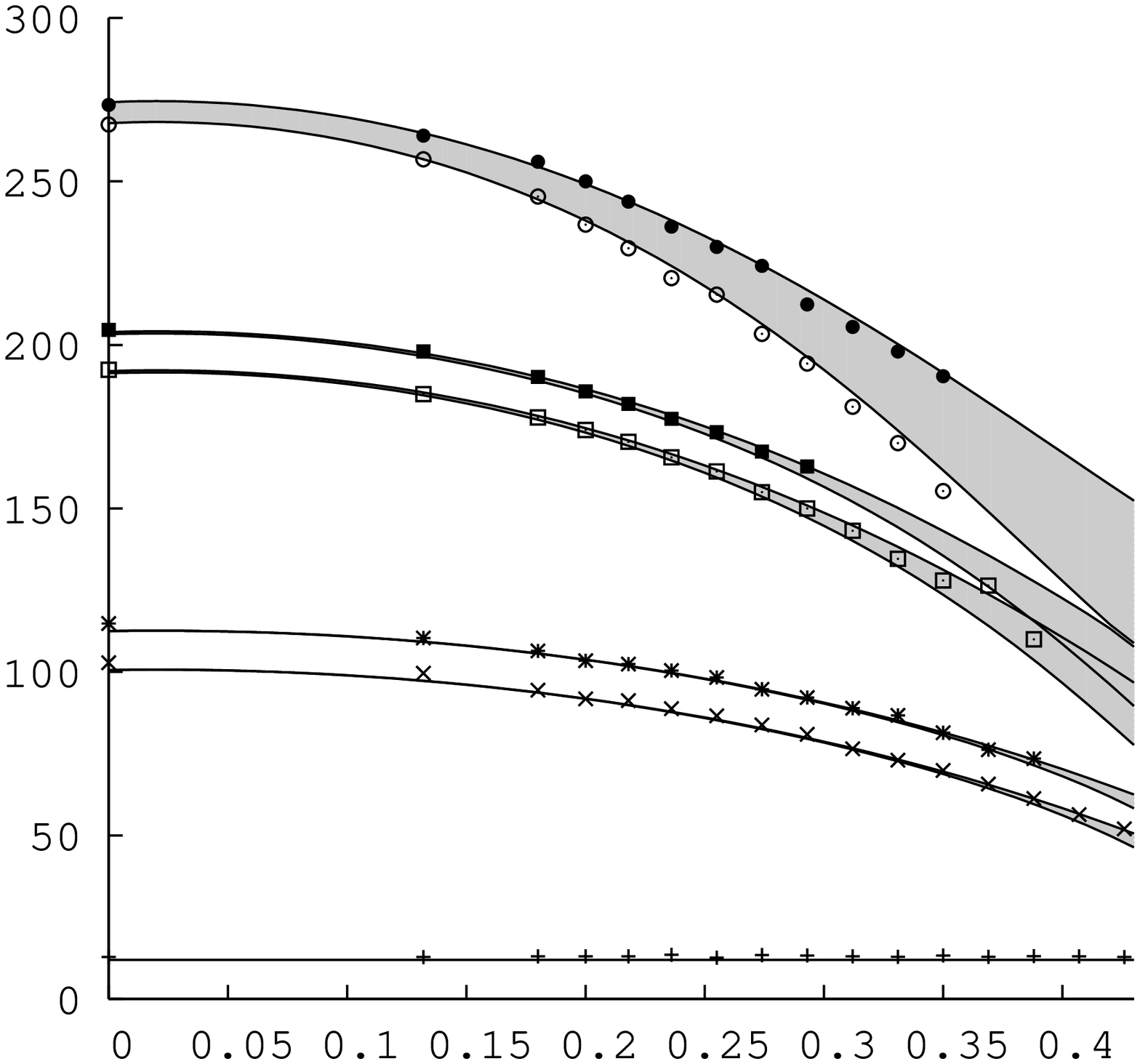}
\put(-115,-202){{${\rm\Phi/\Phi_0}$}}
\put(-224,-110){\rotatebox{90}{{\it V}($\mu$V)}}
\put(-187,-60){$\vert 2,1\rangle$}
\put(-187,-80){$\vert 2,0\rangle$}
\put(-187,-43){$\vert 3(q=e),0\rangle$}
\put(-187,-18){$\vert 3(q=0),0\rangle$}
\put(-187,-169){$\vert 0,1\rangle$}
\put(-187,-110){$\vert 1,1\rangle$}
\put(-187,-135){$\vert 1,0\rangle$}
\put(-80,-60){
\includegraphics[width=0.25\linewidth]{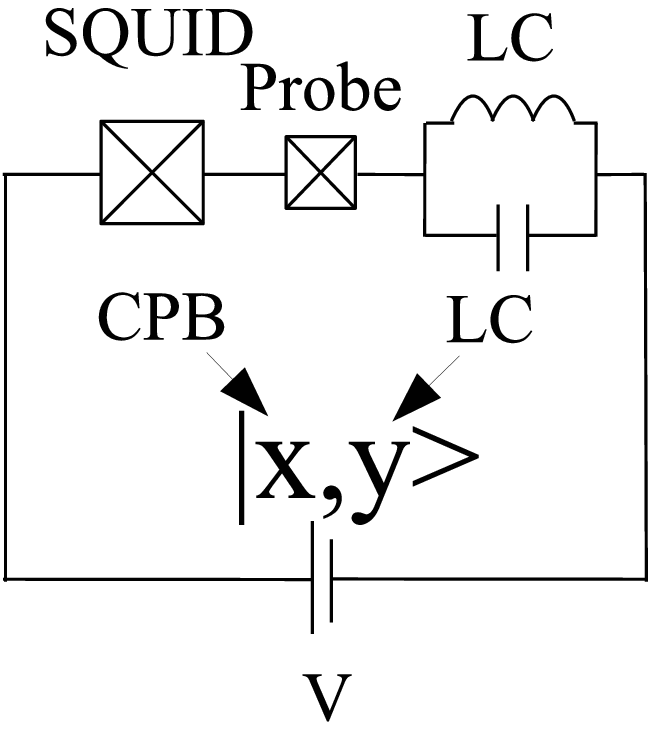}
}
\caption{
The positions of the resonances in the 1-SQUID experiment (data points) compared with those resulting
from the model of a SQUID and an $LC$-oscillator in series with the probe junction (lines).
The resonances have been labelled by the corresponding transitions behind them. For example,
$\vert 1,0\rangle$ means that the resonance occurs due the transition
from the ground state to the product state, in which
the CPB is in its first excited state and the $LC$-oscillator stays in its ground state.
The energy levels of the CPBs are actually bands (shaded), and the resonances due to the band edges $q=0,q=e$ have been plotted explicitly.
The parameters in the numerical modelling are summarized in table~\ref{tab2}.}
\label{paikat1}
\end{center}
\end{figure}

We begin by studying the sample that has the configuration of the asymmetric SCPT (Fig.~\ref{kuvasset}).
The first thing in the fitting procedure is to identify which of the resonances in the $I-V$ characteristics
are coming from the CPB eigenstates, which from spurious fluctuators and which from simultaneous
excitations of both.
The $I-V$ peaks in this "1-SQUID experiment" consist of a set of flux dependent double peaks and several static
resonances, from which the most important is at $V_{LC}\approx 13$ $\mu$V, i.\ e.\ $\omega_0/2\pi\approx 6.3$ GHz,
see Fig.~\ref{paikat1}.
Its second excited state is seen at $2V_{LC}\approx 26$ $\mu$V (not shown in Fig.~\ref{paikat1}) and therefore it is not a
2-state fluctuator.
The resonance is important since it explains the double structure of the first two flux-dependent double peaks:
the lower resonance of each double peak is due to
tunneling of a Cooper pair across the probe and simultaneous excitation of the CPB
and the higher resonance of each double peak is due to tunneling and
simultaneous excitations of the CPB and the $LC$-resonance (a multiphoton transition).
Two observations support this idea.
First, the peak splitting is constant as a function of $E_{J1}(\Phi)=E_{J1}\cos(\Phi/\Phi_0)$
and this constant is the same for the first
and the second double peaks and also equals $V_{LC}$, reflecting the same excitation energy difference $2eV_{LC}$, see Fig.~\ref{paikat1}.
Secondly, when compared to the first peak of a double peak,
the {\it relative} area of the second one is approximately the same for the first two double peaks,
indicating a similar extra factor (matrix element) related to the latter resonance of a double peak.
The theoretical model used in fitting is the same as in section \ref{envi} except that one of the JJs is replaced by an
$LC$-oscillator.

The third double peak is not, however, consistent with similar interpretation, since the peak splitting is smaller
than the previous ones at $\Phi=0$ and increases with increasing $\Phi$. Also,
the two resonances of the double peak have almost equal areas, but
the multiphoton transition to the external $LC$-oscillator should have almost vanishing area and should not even be seen.
The peak splitting can be explained by the band structure of the CPB's third excited state,
assuming that the observed $I-V$ curves are certain averages of the quasicharge-space
and band edges are highlighted due to van-Hove-like singularities.
The observed splitting indeed follows the resonances obtained from band edges, as seen in Fig.~\ref{paikat1}.
Still, the physical reason for the ``escape" of the quasicharge is unknown.
No gate dependence for the positions of the $I-V$ peaks is seen when $E_{J1}>E_c$, supporting the idea of the "running" polarization charge.
It is, interestingly, returned in the limit $E_{J1}<E_c$ at higher voltages as charging effects,
when the gate-dependence of the ground state energy becomes observable.

Indications of the band structure of $\vert 2,0\rangle$ and even $\vert 1,0\rangle$ excitations are seen
in the experiment for $\Phi>0.3\Phi_0$ and $\Phi>0.4\Phi_0$, respectively, but instead of
clear splitting the resonances widen and become fluctuating. This broadening is also expected theoretically, as
shown in Fig.~\ref{paikat1}, and looks like to be caused by random quasicharge fluctuations.
Unfortunately the noise induced by the environment exceeds these current peaks for $\Phi>0.35\Phi_0$ and therefore no
clear evidence of these bands is obtained.

\begin{figure}[tb]
\begin{center}\leavevmode
\includegraphics[width=0.64\linewidth,angle=-90]{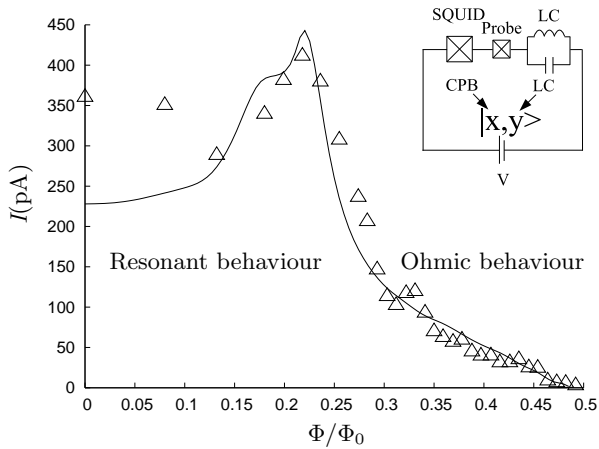}
\put(-115,-165){{${\rm\Phi/\Phi_0}$}}
\put(-228,-83){\rotatebox{90}{{\it I}(pA)}}
\put(-80,-100){{{\rm Ohmic behaviour}}}
\put(-190,-100){{{\rm Resonant behaviour}}}
\put(-75,-70){
\includegraphics[width=0.25\linewidth]{figure4inset.eps}
}
\caption{
The maximum current of the resonance $\vert 1,0\rangle$ in the 1-SQUID experiment (triangles) compared
with that resulting from the model of a SQUID and an $LC$-oscillator in series with the probe junction.
Different from Fig.~\ref{paikat1}, the $LC$-oscillator (``$LC2$" in table \ref{tab2})
describes a small $I-V$ peak seen at a voltage $\approx 90$ $\mu$V (not shown in Fig.~\ref{paikat1}).
In the region $\Phi>0.3\Phi_0$ the state $\vert 1,0\rangle$ suffers from a slow relaxation
and the current is determined by the photon emission to the ohmic low frequency environment.
A resonance occurs nearby $\approx 0.2\Phi_0$ when the states $\vert 1,0\rangle$ and $\vert 0,1\rangle$ are coupled,
the latter having a faster relaxation to its ground state. The current between $\Phi=0$ and $\Phi=0.1\Phi_0$ seems
also to be enhanced, probably due to further entanglement with the environment.
In the calculation of the relaxation rates we have used a perturbation $(C_2/C_{1})Q_{\rm{SQUID}}V_f+0.2Q_{LC}V_{f'}$ (section \ref{envi}).
}
\label{maxvirta}
\end{center}
\end{figure}

Finally, we note that the linewidths of the resonances are similar for the lowest resonances $\sim 8$ $\mu$V,
which can be fitted using the values
$R_0=100$ $\Omega$ and $T= 0.4$ K (leading to $\Delta_{\rm{env}}/2e\sim 3.5$ $\mu$V) and the independently measured value
$E_{J2}=8.5$ $\mu$eV.
The effective temperature is quite high since the temperature of the environment was $\sim 0.1$ K.
The model indicates that the lowest resonances are in the slow relaxation regime,
which is consistent with the observation that the maximum current of the $\vert 1,0\rangle$ resonance decreases when the magnetic flux is
increased, see Fig.~\ref{maxvirta}.

\subsection{The 4-SQUID Experiment}\label{experiments2}

\begin{figure}[tb]
\begin{center}\leavevmode
\includegraphics[width=0.75\linewidth]{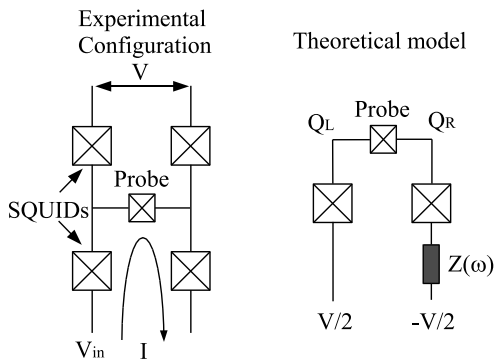}
\caption{Left: Schematic drawing of the 4-SQUID experiment. The
SQUIDs are drawn as JJs. In this situation, the two SQUIDs (at the same side) behave as a single JJ, but with
double the coupling energy $E_J$ and capacitance $C$ compared to the individual SQUIDs. Right: The resulting theoretical model of the system.}
\label{schema3}
\end{center}
\end{figure}

The second sample to be studied consists of four leads, four SQUIDs and a probe junction, see  left side of Fig.~\ref{schema3}.
Since the two SQUIDs on the same side of the probe behave as a SCPT and the phase difference $\theta$ across this component
relaxes to the minimum energy value $\theta=0$ (its classical dynamics is highly damped and no bias is present), the two
SQUIDs behave as a single JJ and one arrives at an equivalent circuit of two JJs and a probe in series connection; the model discussed
in section \ref{envi}.

\begin{figure}[tb]
\begin{center}\leavevmode
\includegraphics[width=0.8\linewidth,angle=-90]{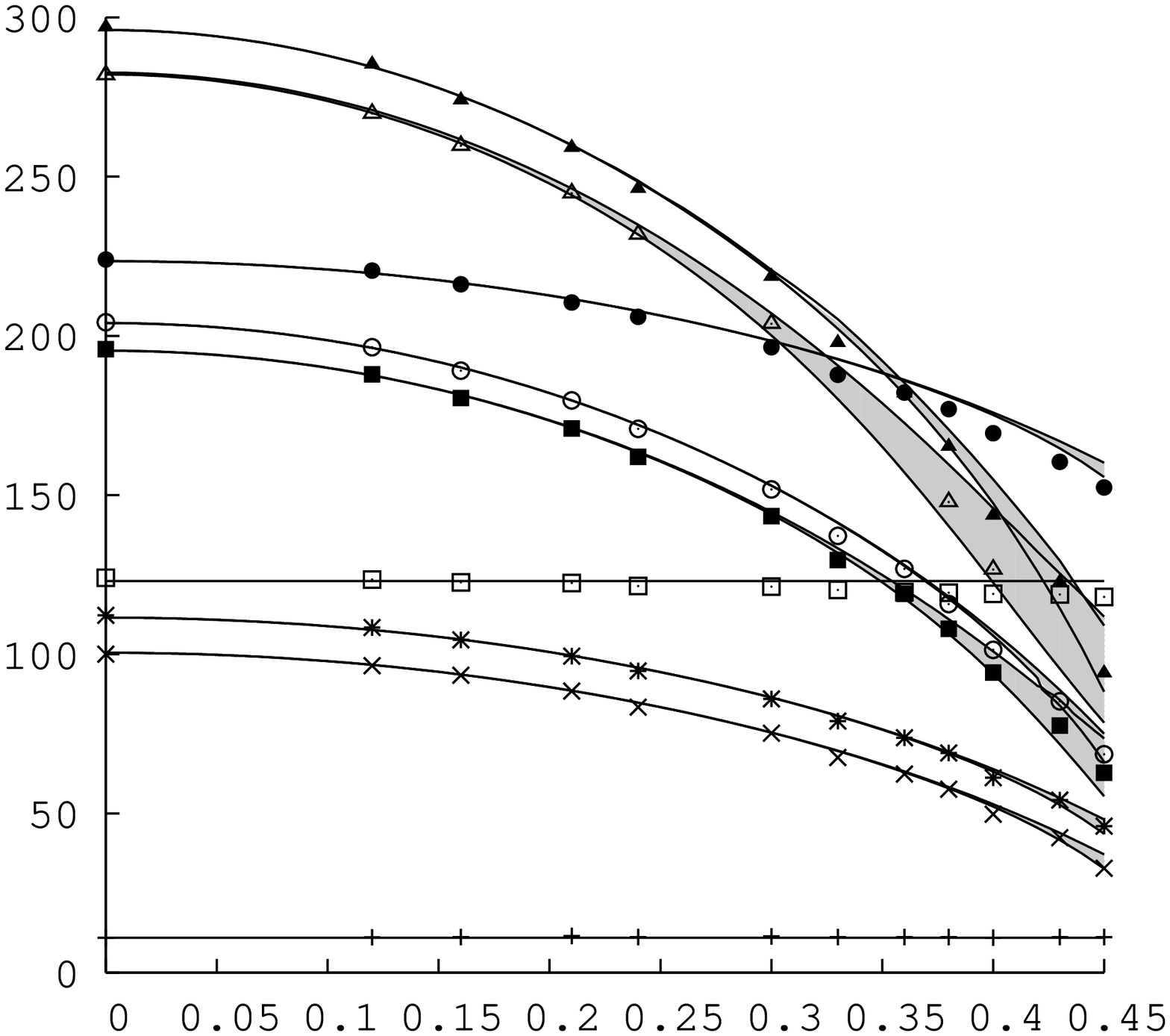}
\put(-113,-202){{${\rm\Phi/\Phi_0}$}}
\put(-224,-110){\rotatebox{90}{{\it V}($\mu$V)}}
\put(-187,-62){$\vert 2^*,0,0\rangle$}
\put(-155,-50){$\vert 1,0,1\rangle$}
\put(-187,-80){$\vert 2,0,0\rangle$}
\put(-187,-33){$\vert 3,0,0\rangle$}
\put(-187,-8){$\vert 3^*,0,0\rangle$}
\put(-187,-169){$\vert 0,1,0\rangle$}
\put(-187,-107){$\vert 0,0,1\rangle$}
\put(-105,-147){$\vert 1,0,0\rangle$}
\put(-92,-125){$\vert 1,1,0\rangle$}
\put(-97,-63){
\includegraphics[width=0.4\linewidth]{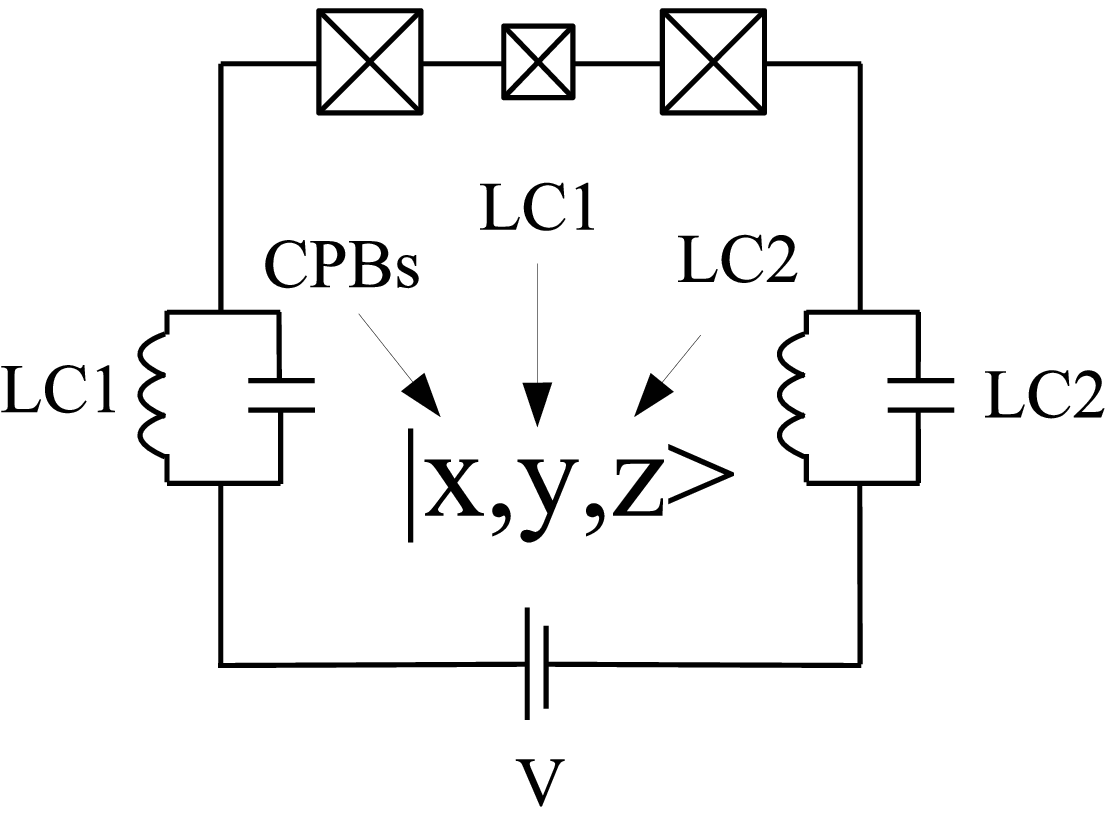}
}
\caption{
The positions of the resonances in the 4-SQUID experiment (data points) compared with
the edges of the energy bands (lines) and the band structure (shaded) calculated from the model
of two SQUIDs and two $LC$-oscillators in series with the probe junction.
The experimental data is based on $I-V$ curves of which two examples are shown in Fig.~\ref{current3}.
Also shown is a schematic diagram of the model circuit and the code used for labelling the states.
The parameters in the numerical model are summarized in table~\ref{tab2}.
}
\label{paikat}
\end{center}
\end{figure}

Also for this sample, the $I-V$ characteristics consist of flux dependent double peaks and a few
static resonances. The noise of the environment in the limit $E_{J1}\sim E_C$ is, however,
much smaller than in the 1-SQUID sample and therefore a more detailed analysis can be done.
Again, the first static resonance is seen at $V_{LC}\approx 11$ $\mu$V i.\ e.\ $\omega_0/2\pi\approx 5.3$ GHz, see Fig.~\ref{paikat},
and its second excited state is seen at $2V_{LC}$ (not shown in Fig.~\ref{paikat}).
We include also a resonance seen at $V_{R}\approx 123$ $\mu$V to the model as another $LC$-oscillator.
The first flux dependent double peak can again be explained a
as a plain excitation of the CPB, and a multiexcitation of the CPB and the (smaller frequency) $LC$-oscillator.
The second double peak, however, is not consistent
with this assumption since the peak splitting is much smaller than $V_{LC}$, $8.5$ $\mu$V, and the relative areas
of the peaks, $1:0.4$, differ essentially from the ones observed for the first double peak, $1:0.13$.
Instead, a better explanation for the second double peak is the
energy level structure of two coupled CPBs in the anharmonic region (section \ref{envi});
the peak splitting occurs due to resonances of the $\vert 2,0,0\rangle$ and $\vert 2^*,0,0\rangle$ excitations,
and using the parameters given in table \ref{tab2} one obtains a peak splitting $8.5$ $\mu$V and an area ratio $1:0.25$.
The resonances corresponding to the multiphoton transitions $\vert 2,1,0\rangle$ and $\vert 2^*,1,0\rangle$ are
also seen as weak peaks in agreement with the model.
The third double peak splitting is then automatically explained as excitations to the states
$\vert 3,0,0\rangle$ and $\vert 3^*,0,0\rangle$, giving the peak splitting  $13.2$ $\mu$eV and area ratio $1:1$.
The experimental values are $15.0$ $\mu$eV and $1:1$.
Note, that in this sample the band structure of the state $\vert 3,0,0\rangle$ is neglible at
$\Phi=0$, and therefore it does not explain the third double peak.
The multiexcitation $\vert 1,0,1\rangle$ is also seen in Figs.~\ref{sjlc8} and \ref{paikat},
justifying generally the multiphoton interpretation.

\begin{figure}[tb]
\begin{center}\leavevmode
\includegraphics[width=0.7\linewidth,angle=-90]{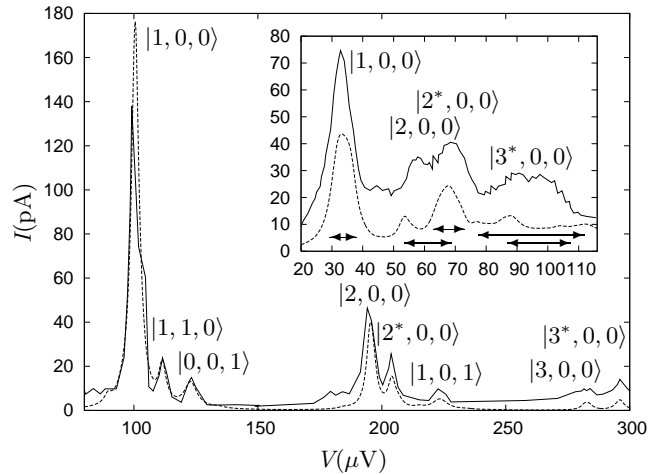}
\put(-130,-180){{\it V}(${\rm\mu V}$)}
\put(-245,-95){\rotatebox{90}{{\it I}(pA)}}
\put(-195,-20){$\vert 1,0,0\rangle$}
\put(-193,-130){$\vert 1,1,0\rangle$}
\put(-183,-143){$\vert 0,0,1\rangle$}
\put(-122,-117){$\vert 2,0,0\rangle$}
\put(-108,-132){$\vert 2^*,0,0\rangle$}
\put(-95,-147){$\vert 1,0,1\rangle$}
\put(-50,-146){$\vert 3,0,0\rangle$}
\put(-45,-133){$\vert 3^*,0,0\rangle$}
\put(-65,-65){$\vert 3^*,0,0\rangle$}
\put(-93,-44){$\vert 2^*,0,0\rangle$}
\put(-103,-54){$\vert 2,0,0\rangle$}
\put(-118,-29){$\vert 1,0,0\rangle$}
\linethickness{0.2pt}
\put(-120,-93){\vector(1,0){5}}
\put(-120,-93){\vector(-1,0){5}}
\put(-88,-95){\vector(1,0){9}}
\put(-88,-95){\vector(-1,0){9}}
\put(-80,-90){\vector(1,0){6}}
\put(-80,-90){\vector(-1,0){6}}
\put(-46,-95){\vector(1,0){12}}
\put(-46,-95){\vector(-1,0){12}}
\put(-49,-92){\vector(1,0){20}}
\put(-49,-92){\vector(-1,0){20}}
\caption{The current across the 4-SQUID sample when $\Phi=0$ (main frame) and $\Phi=0.45\Phi_0$ (inset). The solid line
is the experimental and the dashed line the theoretical $I-V$ curve.
For the theoretical fit we have averaged over all values of $Q_0$ and $Q_0'$ and
the widths of the bands are indicated as arrows. In the inset
one can see that the $\vert 1,0,0\rangle$-resonance has widened and centers at $\sim 32$ $\mu$V.
The $\vert 2^*,0,0\rangle$-resonance lies almost at the same point as the
upper edge of the $\vert 2,0,0\rangle$-resonance, making the overall structure asymmetric.
The $\vert 3^*,0,0\rangle$-resonance lies between $\sim 88$ $\mu$V and $\sim 108$ $\mu$V,
and the $\vert 3,0,0\rangle$-resonance lies between $\sim 78$ $\mu$V and  $\sim 112$ $\mu$V.
In the calculation of the relaxation rates we have used $Q_{\rm{int}}=0.2(Q_1+Q_3+Q_{LC1}+Q_{LC2})$ (section \ref{envi}).
}
\label{current3}
\end{center}
\end{figure}

We conclude that the area ratio and the peak splitting of the first double peak can be fitted by changing the properties
of the external $LC$-oscillator but at the same time the resonance at $V=V_{LC}$ has to be fitted also, whereas the other splittings and areas are determined by the charging energy $E_C$
of the island(s) and the capacitance of the probe junction $C_2$ (section \ref{envi}).
Also adding a slight asymmetry between the two SQUIDs can ``fine tune'' these values and we have used here a 1\% difference.
From Fig.~\ref{current3} one can see that, not only the calculated positions of the resonance peaks, but also the
corresponding areas are quite similar for the model and experiment for this choice of fitting parameters.

\begin{table}[bt]
\begin{tabular}{lccccc}
\hline\hline
Sample& $E_{J1}(\mu$eV) & $E_{J2}(\mu$eV) &  $C_1$(fF) &  $C_2$(fF) & $T$(K)\\ \hline
1-SQUID&  390 (188) &  8.5 (8.5) & 4.6 (5.7) & 0.7 (0.8) & 0.4(0.1) \\
4-SQUID&  483 (544) &   3.6 (3.6) & 6.45 & 0.15 (0.5) &	0.2(0.1)\\
\hline
 & $Z(0)(\Omega)$ & $L_{1}$(nH) &  $L_{2}$(nH) &  $C_{LC1}$(fF) &  $C_{LC2}$(fF) \\ \hline
1-SQUID&  100 &  3.1 & 0.12  & 240 & 100  \\
4-SQUID&  100 &  3.8 & 0.3 & 240 & 24 \\
\hline\hline
\end{tabular}
\caption[vtable]{Parameter values used to fit the experimental $I-V$ curves.
The values resulting from independent measurements~\cite{rene1} are given in parentheses.
We note that the experimental value for $E_{J1}$ in the 1-SQUID experiment,
obtained via a normal state resistance measurement,
is unreliable because it was obtained after the probe was
accidentally broken. When modelling the 1-SQUID sample, we used an asymmetry factor $d=0.12$ for $E_{J1}$, i.\ e.\ 
the Josephson coupling energies of the two JJs inside a SQUID satisfy
$\vert E_{JJ1}-E_{JJ2}\vert /E_{J1}=0.12$.
}
\label{tab2}
\end{table}

It is interesting to study the effect of band structure in this experiment.
Again, a change in the gate voltage did not result in a change of the resonance positions,
even in the region $E_{J1}\sim E_C$. Instead, the resonances originating from SQUIDs widened and
changed from smooth Lorentzians to fluctuating lines.
Therefore, we again assume that the measured $I-V$ curves are a result of some kind of
averaging over the quasicharge space, which now is two-dimensional because of two quasicharges $Q_0$ and $Q_0'$.
The theoretical and experimental $I-V$ curves are compared at $\Phi=0.45\Phi_0$ in  the inset of Fig.~\ref{current3}.
One can see that they compare fairly well, even though the peak heights for show little disagreement.
In this sample the band structure does not lead to strong peak splitting in contrast to the 1-SQUID sample.
The reason for this is that the van Hove singularities in two dimensional
quasicharge space are weaker than in one dimension.
This is not the case for single excitations $\vert 1,0,0\rangle$, $\vert 2,0,0\rangle$ and $\vert 3,0,0\rangle$,
and indeed the band edges of the latter two are seen as separate peaks in the theoretical $I-V$ curve.
We have used uniform quasicharge distribution, which is the simplest guess as the physics of the average processing
are unknown.

\begin{figure}[tb]
\begin{center}\leavevmode
\includegraphics[width=\linewidth]{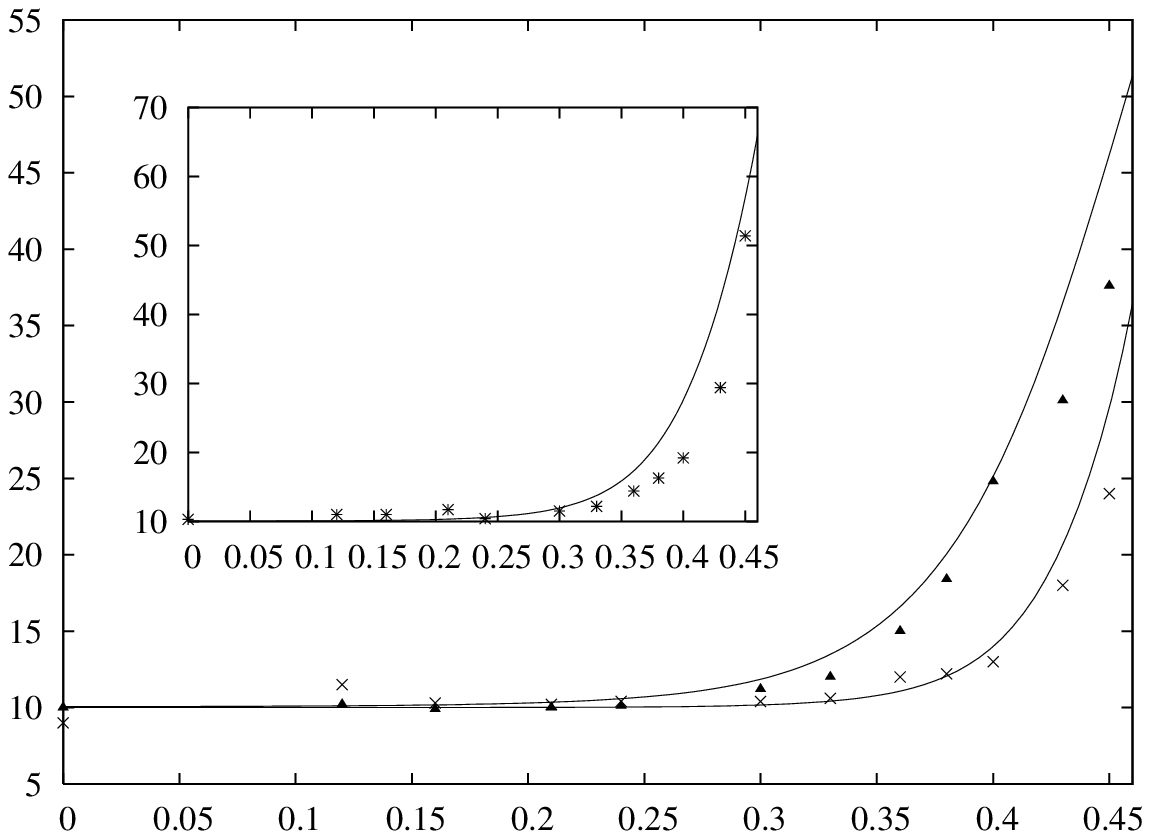}
\put(-245,78){\rotatebox{90}{$\Delta$($\mu$eV)}}
\put(-128,-7){${\rm\Phi/\Phi_0}$}
\put(-160,130){$\vert 3^*,0,0\rangle$}
\put(-52,115){$\vert 2,0,0\rangle$}
\put(-47,25){$\vert 2^*,0,0\rangle$}
\caption{Experimental linewidths
(data points) compared with theoretical linewidths (lines)
for three of the resonances as a function of the magnetic flux.
The theoretical values are obtained by adding a constant term $10$ $\mu$eV to the widths of the bands.
}\label{leveys3}\end{center}\end{figure}

The experimental and theoretical peak broadening are compared in Fig.~\ref{leveys3} for three of the transitions.
The theoretical width is obtained by summing up the peak width at $\Phi=0$, which for this choice of parameters is
$\approx 10$ $\mu$eV, and the increase due to broadening of the bands.
There is good agreement between the theory and experiment. For example,
the $\vert 2,0,0\rangle$-resonance broadens faster than the $\vert 2^*,0,0\rangle$-resonance
which is consistent with the theoretical model.
The width of the $\vert 3,0,0\rangle$-resonance is not analyzed since it rapidly becomes unobservable due to strong
broadening.

\section{Conclusion}\label{conclusion}

We have carried out a theoretical study of Cooper pair tunneling across a voltage biased
asymmetric SCPT and a system consisting of three JJs in series, where the middle one acts as a probe, and applied the models in analyzing
the experimental findings of Ref.~\onlinecite{rene1}.
The treatment of the problem was done in the weak coupling regime, where the Cooper pairs tunnel incoherently
across the probe, and was based on the idea of extending the well known $P(E)$-theory
into the regime where the anharmonicity and band structure are taken into account.
We pointed out, that the nature of the tunneling
across the probe turns from incoherent to coherent when the golden rule tunneling times exceed the
relaxation times induced by the dissipative environment. Furthermore,
we discussed that a simple master equation correction to the population of the eigenstates in
the incoherent calculation leads to a good approximation for the current for arbitrary values of the voltage 
and for different flux values.

In the last part of this paper we showed that
a detailed theoretical understanding of experimental data can be achieved. In particular, the
multiphoton processes between
different mesoscopic elements and spurious $LC$-resonators as well as the band structure of the Josephson junction
can be probed by a small Josephson junction coupled to SQUID(s). Especially, the detection of energy bands of higher excited states is
confirmed by the fact that the observed widening of the resonances was in good accordance with the linewidths
obtained from the model.

\acknowledgments

This work was financially supported by the National Graduate School in Material Physics, Finnish Academy of Science
and Letters (Vilho, Yrj\"o and Kalle V\"ais\"al\"a Foundation) and the Academy of Finland (National Center of Excellence Programme).


\end{document}